# Frictional Transition from Superlubric Islands to Pinned Monolayers

Matteo Pierno[1], Lorenzo Bruschi[2] and Giampaolo Mistura[1]

Guido Paolicelli[3], Alessandro di Bona[3] and Sergio Valeri[4,3]

Roberto Guerra[5,6], Andrea Vanossi[5,6] and Erio Tosatti[5,6,7]

[1]Dipartimento di Fisica e Astronomia G. Galilei, Università di Padova, via Marzolo 8, 35131 Padova, Italy

[2] CNISM Unità di Padova, via Marzolo 8, 35131 Padova, Italy

[3] CNR, Istituto Nanoscienze - Centro S3, via Campi 213/a, 41125 Modena, Italy

[4]Dipartimento di Scienze Fisiche Informatiche e Matematiche, Università di Modena e Reggio Emilia, via Campi 213/a, 41125 Modena, Italy

[5]International School for Advanced Studies (SISSA), via Bonomea 265, 34136 Trieste, Italy

[6]CNR-IOM Democritos, via Bonomea 265, 34136 Trieste, Italy

[7]International Centre for Theoretical Physics (ICTP), Strada Costiera 11, 34151 Trieste, Italy

**The inertial sliding of physisorbed submonolayer islands on crystal surfaces contains unexpected information on the exceptionally incommensurate smooth sliding state associated with superlubricity and on the mechanisms of its disappearance. Here we show in a joint quartz crystal microbalance (QCM) and molecular dynamics (MD) simulation case study of Xe on Cu(111) how superlubricity emerges in the large size limit of naturally incommensurate Xe islands. Theory also predicts, as coverage approaches a full monolayer, an abrupt adhesion-driven 2D density compression of order several %, implying here a hysteretic jump from superlubric free islands to a pressurized $\sqrt{3}\times\sqrt{3}$ commensurate immobile monolayer. This scenario is fully supported by QCM data showing remarkably large slip times at increasing submonolayer coverages, signaling superlubricity, followed by a dramatic drop to zero for the dense commensurate monolayer. Careful analysis of this variety of island sliding phenomena will be mandatory in future applications of friction of crystal-adsorbate interfaces.**

Systems achieving low values of dry sliding friction are of great physical and potentially technological interest[1-4]. Superlubricity, the vanishing of static friction with consequent ultra-low dynamic friction taking place typically between crystal faces that are sufficiently hard and mutually incommensurate [5,6], is experimentally rare, and has been demonstrated or implied in a relatively small number of cases, such as telescopic sliding among carbon nanotubes[7,8], sliding graphite flakes on a graphite substrate[9-11], cluster nanomanipulation[12,13] and sliding colloidal layers[14,15]. It is mandatory to increase our understanding and, also in view of potential nanotechnology applications, to examine new and more generic systems beyond these.

Submonolayer islands of rare gas atoms adsorbed on crystal surfaces offer an excellent platform to address friction at crystalline interfaces. Despite much experimental[16-22] and theoretical[23-27] work, superlubricity is an aspect which is still poorly explored precisely in these systems. Adsorbate phase diagrams versus coverage $\theta$ are well known to display, in the submonolayer range ($0 < \theta < 1$) and at low temperatures, phase separated 2D solid islands, usually incommensurate with the surface lattice, coexisting with the 2D adatom vapor[28,29]. In the Quartz Crystal Microbalance (QCM), the inertial sliding friction of these islands is measured by the inverse of the slip time $\tau_s = (1/4\pi) [\delta(Q^{-1})/\delta f]$, the ratio of the adsorbate-induced change of inverse quality factor over the respective change of the substrate oscillations frequency[30]. The peak of the inertial force acting on an island deposited on the QCM is $F_{in} = \rho_{isl} S A (2\pi f)^2$ (where $\rho_{isl}$ is the 2D density at the center of an adsorbed island of area $S$, while $A$ and $f$ are the oscillation amplitude and frequency, respectively) equals the viscous frictional force of $F_{visc} = M\mathbf{v}/\tau_s$, (island mass $M$ and speed v). Superlubricity should thus indirectly show up as an unusually large slip time. For over two decades, QCM work has shown that physisorbed atoms or molecules condense and generally slide above some submonolayer coverage $\theta_{sf}$, where $\tau_s$ may reach typical values of hundreds of ps to a ns. These results[17,19,30] and the corresponding pioneering atomistic simulations[23,31] have brought much valuable initial information about the temperature and system dependence of inertial friction. So far however, crucial aspects specifically addressing the island structure of the adsorbate, the edge originated pinning and, especially, the change of commensurability and superlubricity with coverage, issues that are in our view important to nanofriction, have not yet come into scrutiny.

Here we present a joint experimental and theoretical study of the sliding of adsorbate islands on a crystalline substrate revealing surprising information about the exceptionally easy sliding suggestive of superlubricity, about its limiting factors caused by edges and defects, eventually its spontaneous demise at full coverage. Our chosen example is physisorbed Xe on Cu(111), a system whose phase diagram is, along with other rare gas adsorbates on graphite and metal surfaces, well studied[28]. Between about 50 and 90 K, Xe monolayers condense on Cu(111) as a commensurate √3×√3 2D solid. We conventionally designate this as unit coverage $\theta$=1, characterized by a density $\rho$=$\rho_0$ = 2/(√3 $a_{Cu}^2$) where $a_{Cu} = \sqrt{3}d_{Cu-Cu}$ = 0.441 nm is the commensurate adatom spacing. Low-energy electron diffraction at 50 K locates the Xe atoms on top of surface Cu atoms[32], whose planar distance $a_0$ is close to the Xe-Xe spacing in bulk Xe, $a_{Xe}$ = 0.439 nm. At lower temperatures the full Xe monolayer is known from surface extended X-ray measurements to shrink into an "overdense" ($\rho > \rho_0$) incommensurate structure, reaching √3×√3 commensurability only at 50 K upon thermal expansion[33]. Conversely, the 2D atom density in Xe monatomic *islands*, which coexist with the adatom 2D vapor at submonolayer coverage, is not specifically known, although often assumed to be equal $\rho_0$. Our results actually show that the 2D crystalline Xe islands ($\theta$<1) are slightly "underdense" ($\rho_{isl} < \rho_0$) and increasingly incommensurate with thermal expansion, reaching a 2D density 4% below $\rho_0$ near 50 K. In this incommensurate state, the 2D lattice inside the Xe islands should slide superlubrically over the Cu(111) substrate, as expected for a "hard" slider. Indeed, even though the Xe-Xe attraction $V_{Xe-Xe}$~20 meV is an order of magnitude smaller than the Xe-Cu(111) adhesion energy $E_a$ ~190 meV[28], it is an order of magnitude larger, and thus harder, than the weak Cu(111) surface corrugation $E_c$ ~ 1-2 meV[34]. In addition, we find that the large Xe/Cu adhesion entails another important consequence whose tribological impact has not generally been described. At monolayer completion, where the 2D adatom gas disappears, a positive 2D (spreading) pressure suddenly builds up as extra adatoms strive to enter the first layer and benefit of the substrate attraction, as opposed to forming a second layer where attraction is much smaller. This process, for example, is clearly revealed in QCM data for Xe/Ag(111) by Krim's group[17], where it gave rise to a Xe density increase of about 5 % upon monolayer formation. Ideally, as the submonolayer coverage grows, this spontaneous density increase process should start at $\theta_c \sim \rho_{isl}/\rho_0 < 1$ and continue until limited by either the buildup of 2D pressure, or by a strong accidental commensurability with the substrate, whichever comes first as $\theta$ grows beyond $\theta_c$. If corrugation, commensurability, and entropic effects are ignored, and assuming for simplicity a first-neighbor Xe-Xe attraction $-V$, the potential energy density change upon a monolayer density increase from $\rho$ to $\rho+\delta\rho$ is roughly estimated as

$$\delta E = \rho(-E_a + 3V)(\delta\rho/\rho) + (1/2)(\lambda + \mu)(\delta\rho/\rho)^2 \qquad (1)$$

which is minimal when

$$(\delta\rho/\rho) = \rho(E_a - 3V)/(\lambda+\mu) = (E_a - 3V)/mv_L^2 \qquad (2)$$

where $\lambda$ and $\mu$ are the Xe monolayer Lamé coefficients, $m$ is the mass of a Xe atom and $v_L$ is the monolayer longitudinal sound velocity. With parameters appropriate for the Xe monolayer and $v_L$ =1.3 Km/s), this yields $(\delta\rho/\rho) \sim 6\%$, close to the experimental compression of Xe/Ag(111). We note that a compression of this magnitude would amount to several Kbar in bulk Xe. In the present case of Xe/Cu(111), and unlike Xe/Ag(111), the 2D density upward jump from $\rho_{isl}$ is arrested to $\rho_0$ by $\sqrt{3}\times\sqrt{3}$ commensurability, and we estimate its amount to about 4 % . The gist of these preliminary theoretical considerations is that, near 50 K, submonolayer Xe islands *must* be incommensurate, and most likely superlubric, whereas the full monolayer is $\sqrt{3}\times\sqrt{3}$ commensurate, and probably pinned.

**QCM measurements** The friction of Xe monolayers has been measured with a quartz crystal microbalance (QCM). The microbalance consists of a AT cut quartz disk whose principal faces are optically polished and covered by a gold keyhole electrode commercially evaporated on one face and a copper one on the other face (see Fig. 1). The QCM was driven at its fundamental mode with resonance frequency $f_{res} \sim 5$ MHz by using a frequency-modulation (FM) technique. An AC voltage $V_D$ is applied across the two electrodes at a frequency equal to that of its mechanical resonance and drives the two parallel faces of the quartz plate in an oscillating, transverse shear motion. Varying $V_D$ changes the power dissipated in the quartz and the amplitude A of the lateral oscillations of the electrodes. The latter quantity is calculated from the formula $A = 1.4 Q_S V_D$, where A is measured in pm and $V_D$ is the peak driving voltage in V[35]. The top graph in Fig. 1 shows a family of resonance curves measured in vacuum and at T = 48 K for different A. In the horizontal axis the frequency of the generator f is normalized to $f_{res}$, while the vertical axis shows the corresponding amplified voltage $V_{QCM}$ normalized to the peak value $V_{res}$. No variation in the resonance curve is detected within the amplitude range investigated. The continuous line is a nonlinear least-squares fit to the data that yields a quality factor of the quartz Q=220000[20]. The condensation of a film on the electrodes is signaled by a decrease of the resonance frequency $f_{res}$. Any dissipation taking place at the solid-film interface is instead detected by a decrease in the corresponding resonance amplitude $V_{res}$[36].

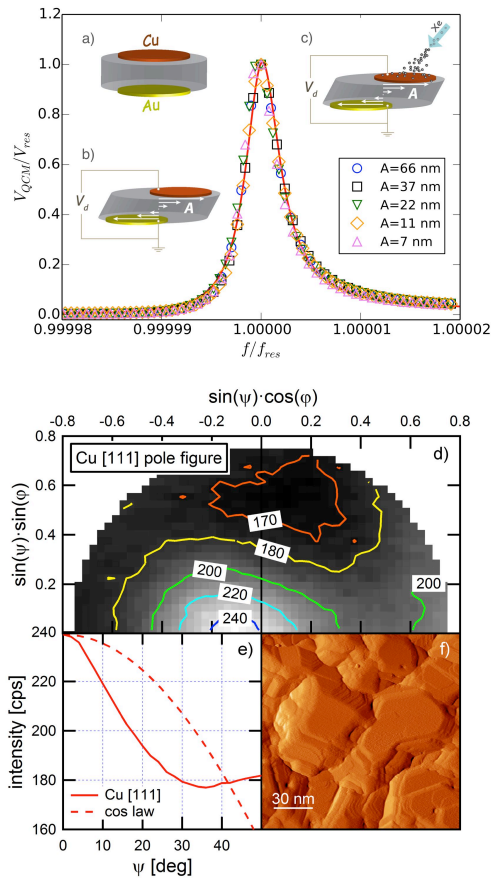

**Figure 1 – The quartz crystal microbalance and characterization of the Cu(111) electrode.** Top: normalized resonance curves of the QCM measured in vacuum and at $T$=48 K for different oscillating amplitudes $A$. The red line is a fit to the data for $A$=7nm. $f_{res}$~5 MHz represents the series resonance of the quartz crystal. A sketch of the QCM with the Cu and Au electrodes is shown in a). Inset b) indicates the shear motion of the QCM at resonance: arrows represent the lateral displacements. Inset c) shows Xe gas dosing on the QCM. Bottom: d) Pole figure (stereographic plot) of the Cu [111] peak. The contour plots labels unit is cps; ψ (sample tilt angle) is varied in the 0° - 50° range while φ (sample rotation angle) is varied in the 0° - 180° range. e) Continuous line: φ-averaged Cu [111] intensity as a function of ψ; dashed line: $cos(\psi)$ law normalized to Cu [111] intensity at ψ = 0°; f) STM derivative image of the Cu film. The image size is 150 × 150 nm.

The Cu(111) electrode was prepared by depositing on the other bare quartz face a Cu 30 nm / Cr 10 nm bilayer at room temperature in UHV conditions using Knudsen effusion sources. The Cr buffer layer was used to promote adhesion between the quartz substrate and the Cu film. The deposition rate was 2.2 Å/s and 4.7 Å/s for Cr and Cu, respectively. Prior to deposition, the QCM was heated for 30 minutes to 250 °C to remove the condensed impurities. After the deposition, surface cleanliness of Cu film was checked by XPS spectroscopy (see Supplementary Material). Fig. 1(d) shows the XRD intensity of the Cu [111] reflection as a function of the sample orientation (pole figure). A well-defined peak close to 0° tilt angle indicates that most of the Cu grains are oriented with the [111] crystal axis perpendicular to the sample surface. No other preferred orientations are detected. Geometrical effects due to the limited size of the sample with respect to the X-ray beam give rise to a broad peak at $\psi=0°$, with a $\cos(\psi)$ dependence. Fig. 1(e) shows the [111] reflection intensity, averaged over the sample rotation angle $\varphi$, as a function of the sample tilt angle $\psi$. To discriminate between preferential orientation and finite sample size effects, in Fig. 1(e) a $\cos(\psi)$ law normalized to the maximum intensity of the peak is also shown. The measured intensity peak is clearly sharper than $\cos(\psi)$, indicating that the $\psi$ dependence of the [111] intensity is mostly due to the preferential orientation of the Cu grains. Fig 1(f) shows a STM derivative image of the sample surface taken in situ right after the deposition. Large flat grains 40-50 nm in lateral size are clearly visible. The typical area of the (111) terraces was $A_0 \sim 2.5 \ 10^3$ nm$^2$.

Xe was condensed directly onto the Cu(111) electrode of the QCM at temperatures comprised between 47 and 49 K. Lower temperatures could not be reached due to the poor thermal coupling to the cold head of the cryocooler, higher values were limited by the evaporation of the Xe monolayer[22]. Within this very narrow temperature interval, no systematic and reproducible variations attributable to *T* were observed. Between consecutive deposition scans, the QCM was heated to about 60 K to guarantee the full evaporation of Xe and the thermal annealing of the microbalance[37].

Fig. 2 shows the measured slip time $\tau_s$ of Xe at *T*= 47 K with a moderate oscillating amplitude *A* ~7 nm of the Cu electrode. The coverage is deduced from the frequency shift assuming for the monolayer an areal density $\rho_0 = 5.93$ atoms/nm$^2$ corresponding to the completion of the $\sqrt{3}\times\sqrt{3}$ commensurate solid phase, equivalent to a frequency shift of 7.3 Hz. Besides some initial pinning ($\tau_s =0$) at the lowest coverages $\theta < 0.05$ where Xe is known to condense at steps and defects[38], the data show depinning with a rapid increase of $\tau_s$, reaching peak values up to 4 ns, nearly an order of magnitude larger than the slip times measured with Xe on gold and on graphene at the same temperature[22]. This slipperiness of Xe islands on Cu(111) is all the more puzzling because it runs contrary to the pinning for the √3×√3 commensurability so far supposed to be in place at 47 K from literature[33]. The large, fast rising, submonolayer slip time is a dominant feature in Fig.1, which we will now qualify as evidence of incommensurability and superlubricity. The second, even more unusual feature is the sudden slip time collapse near $\theta \sim 1$, also exhibiting a mysterious variability between different experiments. Both will be shown to signify an abrupt increase of density leading to a compressed √3×√3 monolayer.

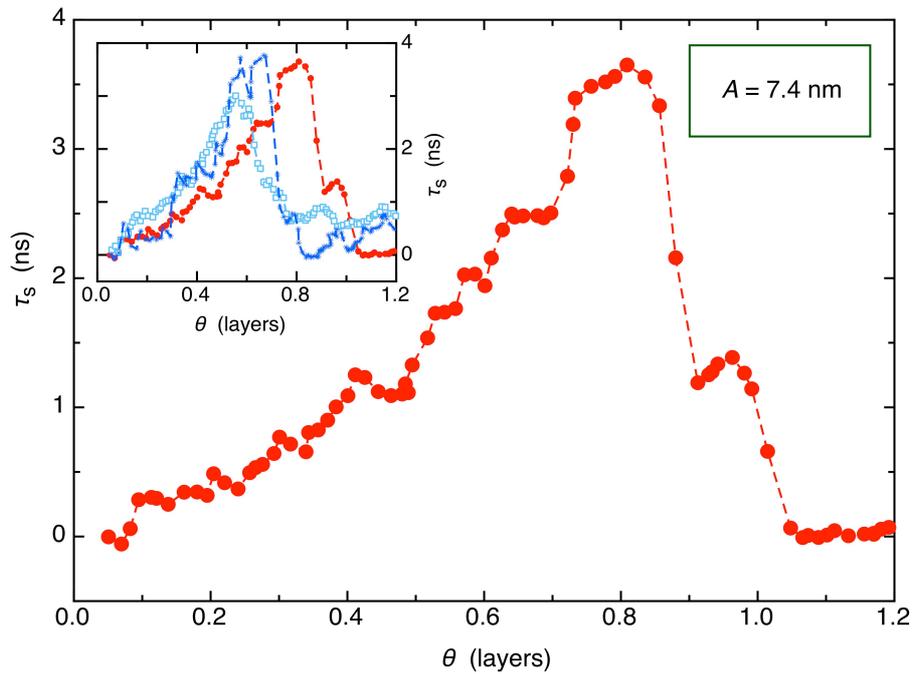

**Figure 2 – Slip time of Xe on Cu(111) as a function of film coverage.** The scan was taken at $T$= 47 K with $f_{res} \sim 5$ MHz and at an oscillating amplitude of the Cu electrode $A$ =7.4 nm. Inset: scans of Xe on Cu(111) taken for different Xe depositions on the same substrate at the same $A$ and temperatures comprised between 47 and 49 K. Note the sharp drop occurring at coverage near a full monolayer, albeit with large fluctuations.

**Theory and MD simulations of sliding islands** The physics behind these experimental results can be directly addressed by frictional molecular dynamics simulations. The power $P$ dissipated by a hard crystalline island of area $S$ sliding on a crystal surface under a uniform force $F$ is the sum $P = P_b + P_e$ of an intrinsic "bulk" term $P_b$ (the friction of an equal portion of infinite adsorbate of same 2D density) and of extrinsic or defect terms $P_e$, representing the correction due to the island finite size, and to substrate defects. The contribution to $P_e$ due to substrate imperfections and defects depends on the oscillation amplitude, and is reduced by smaller oscillation amplitudes, when fewer defects are overlapped by the moving island (see Supplementary Information). The residual contribution to $P_e$, present even on defect-free terraces and for all amplitudes, is the friction caused by the island finite size, that can be conventionally designated as an "edge" contribution[27] . Quite generally, $P_e$ presents a different area dependence from $P_b$, namely $P_b \sim S$ against $P_e \sim S^\gamma$, where $\gamma < ½$ depending on the nature of pinning centers[27].

The force dependence of the two terms also differs. Extrinsic defects and/or island edges imply a small but nonzero static friction force $F_{se}$, so that $P_e$ will vanish for either $F < F_{se}$ due to pinning and for $F >> F_{se}$ where the sliding becomes asymptotically free. The bulk, intrinsic frictional power $P_b$ depends strongly on commensurability of the island 2D lattice structure with the crystal surface lattice. Hard incommensurate islands have zero bulk static friction, hence will slide superlubrically, that is with a relatively small kinetic friction, growing viscous-like with force, $F^\nu$ with $\nu \sim 1$.

This bulk friction of a superlubric slider, negligible at low speed sliding such as µm/s typical of AFM[13], becomes accessible in QCM, where peak speeds are many orders of magnitude higher (here v$\sim \omega A \sim$ 0.23 m/s). Commensurate systems are, on the other hand, pinned by static friction $F_{sb}$, so that $P_b$ is zero until the force reaches a large depinning force $F \sim F_{sb}$, at which point, as shown e.g., by colloidal simulations[15, 39], frictional dissipation has its peak.

In our model, the Cu(111) substrate is treated as a fixed and rigid triangular lattice, exerting on the mobile Xe adatoms an average attractive potential $-E_a$ = -190 meV, and a corrugation 1 meV between the Cu on-top site (energy minimum for a Xe adatom), and the Cu hollow site (energy maximum). Each Xe adatom is thus subject to the overall potential $V_{Xe-Xe}+V_{Xe-Cu}$. The Xe-Xe interaction is modeled by a regular Lennard-Jones 12-6 potential, with $\epsilon$ = 20 meV and $\sigma$ = 3.98 Å. Smaller corrections due to three-body forces as well as substrate-induced modifications of this two-body force are ignored. The Xe-Cu interaction is modeled by the modified Morse potential:

$$V_{Xe-Cu} = \alpha(x,y)\left(e^{-2\beta(z-z_0)} - 2e^{\beta(z-z_0)}\right) \tag{3}$$

where $z_0$ = 3.6 Å [40, 41]. We define the modulating function, normalized to span the interval from 0 (top sites) to 1 (hollow sites)

$$M(x,y) = \frac{2}{3} - \frac{4}{9}\cos\left(\frac{2\pi x}{b}\right)\cos\left(\frac{2\pi y}{\sqrt{3}b}\right) - \frac{2}{9}\cos\left(\frac{4\pi y}{\sqrt{3}b}\right), \tag{4}$$

where the constant $b = 3.61/\sqrt{2}$ is the nearest neighbor distance of surface Cu atoms. The Morse potential energy parameter is given by $\alpha(x,y) = -E_a + M(x,y) E_c$. The inverse length $\beta$ in (3) is obtained by equating the second derivative of the potential to the experimental spring-constant,

$$V''(z_0) = 2\alpha^{top}\beta^2 = \omega^2 m, \tag{5}$$

where $m$ is the atomic mass of Xenon. With a perpendicular vibration energy of $\hbar\omega \simeq 2.8$ meV [40], we obtain $\beta = 0.8$ Å$^{-1}$.

The equations of motion are integrated using a velocity-Verlet algorithm, coupled to a Langevin thermostat with a damping coefficient $\gamma=0.1$ ps$^{-1}$, a damping whose value is not critical, and which is not applied to the translational degrees of freedom of the Xe island center of mass (CM). Islands are obtained by a circular cutting of a Xe monolayer, whose radius determines the island size. The so-formed islands are deposited on the Cu substrate with a random orientation angle, as expected to occur experimentally and not critical to the results. The simulation protocol for slip-time calculation starts with heating the system at 48 K for 100 ps. Then, the Xe CM velocity along x-axis is set at $v_i$=100 m/s, and the simulation evolves until motion stops. The slowdown is very well fit by an exponential, indicating a purely viscous friction. The slip-time is extracted, after skipping the initial transient, by an exponential fit of the form $v(t) = v_i e^{-t/\tau_s}$, as shown in Fig. 3 for an island of diameter ~60 nm.

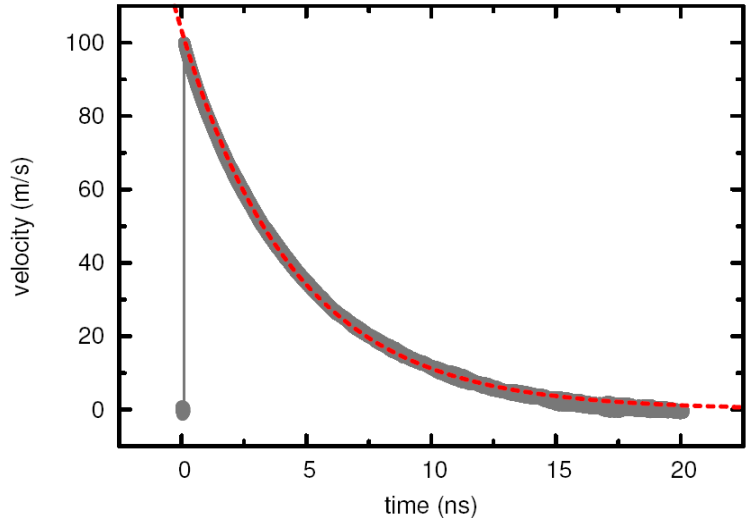

**Figure 3. - Spontaneous frictional slowdown of a 60 nm circular island**. The island of density $\rho_{isl}/\rho_0 = 0.96$ was initially kicked at large speed and T=48 K and then let free to move without thermostating. The grey line is obtained by superposition of five simulations. The excellent exponential fit confirms the viscous sliding of the island.

The excellent exponential fit confirms that the island sliding is indeed viscous, with a slip time $\tau_s$ as large as 5 ns, directly comparable with the experimental values of Fig.2. Moreover, even if no extrinsic defects were included, the slip time obtained still varied with the island area $S$, well fit by $\tau_s^{-1} = (a + b\, S^{\gamma-1})$, with $\gamma \sim \frac{1}{4}$ as shown in Fig. 4. This sublinear term exponent is the same found for sliding clusters[13] and similar to that of a recent study of adsorbate static friction, where it was due to the island finite size[27]. Under the reasonable assumption that increasing coverage corresponds to increased average island size, eventually reaching at full coverage ($\theta \sim 1$) the size of the largest Cu(111) terraces (about 50 nm), the increase of the experimental slip-time with coverage of Fig.2 can be attributed to the progressively decreasing role of edges[27]. The intrinsic, defect free slip time asymptotically reached in the large size limit is that dictated by the ideal Xe lattice, which is incommensurate, hard, and superlubrically sliding with a friction growing linearly with speed. We conclude that the unusually large slip times at large submonolayer coverages signifies precisely that Xe islands sliding on Cu(111) are asymptotically superlubric.

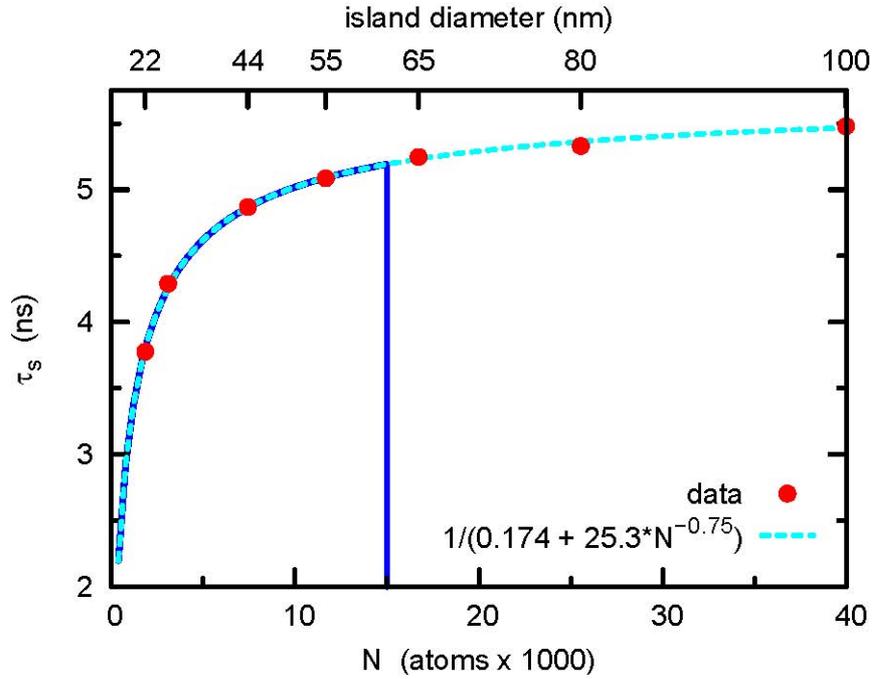

**Figure 4. - Theoretical slip time from sliding simulations for incommensurate Xe islands of growing size, on a perfectly periodic potential representing Cu(111).** The fit (dashed line) shows that the size-controlled defect friction (here due to the island edge) gives way to very long slip times arising from bulk superlubricity in the large size limit ($\tau_s^{bulk} \approx 5.75$ ns). The vertical line marks full monolayer coverage for an estimated experimental Cu(111) terrace size of 60 nm. When islands reach this size, we expect a spontaneous density increase with sudden slip time collapse due to $\sqrt{3} \times \sqrt{3}$ commensurabiliy.

The second striking experimental feature is the sudden drop of slip time near monolayer coverage. The physics behind that also emerges from simulation, where additional adatoms added near full coverage $\theta \sim 1$ get spontaneously incorporated in the first monolayer rather than forming a second layer. As trial simulations also confirms, the extra compressional energy cost implied by this incorporation is overcompensated by the adhesive energy gain, in agreement with Eq. (2) above. The resulting 4% growth of 2D density at monolayer completion, not far from the full theoretical 6%, is arrested by the intervening, and accidental, exact $\sqrt{3} \times \sqrt{3}$ commensurability, well established experimentally[32, 34]. Due to that, the slip time of Xe/Cu(111) falls (unlike that of e.g., Xe/Ag(111), which remains incommensurate after densification[17]), as seen in the experimental Fig.2 and indicated in the theoretical Fig. 4.

The 2D density jump which destroys superlubricity for increasing adsorbate coverage near one monolayer is a sudden, first-order event. As such, it is expected to occur with hysteresis, which implies a difference between atom addition and atom removal, as well as occasional differences between one compressional event and another. As shown in the inset of Fig 2, the Xe coverage at which the sudden slip time drop occurs is experimentally rather erratic, in agreement with this expectation. Experimental verification of this hysteresis is difficult because of the negligible pressure of the bulk vapor in equilibrium with the film, which makes it impossible to decrease the Xe coverage by pumping gas out at the temperature of the scan. Simulated insertion/extraction of a Xe atom into/out of a full underdense monolayer ($\rho/\rho_0 = 0.96$) yields very asymmetric energy evolutions, actually ending with highly defected, poorly reproducible states. That result supports hysteresis and moreover suggests the explanation for the randomness in the slip time drop observed in the real process of Fig. 2, probably due to a relatively long, statistically distributed time needed for the intra-monolayer defects to heal away during the spontaneous compression process.

The temperature behavior, before closing, is one important parameter which we cannot vary in our experiment, but which is worth commenting upon. The compressed Xe/Cu(111) monolayer has a much stronger thermal expansion that bulk Xe, which causes it to evolve from incommensurate and slightly overdense between 20 to 45 K[32] to $\sqrt{3}\times\sqrt{3}$ commensurate at 50 K[31,32] and 60 K[33], pushing it incommensurate again, now slightly underdense, at higher temperatures (a fine feature hard to pick up by pioneering LEED studies[42]). Incommensurability and superlubricity at 77 K are strongly suggested by the exceedingly long slip times of order 20 ns observed by Coffey and Krim[18].

**From superlubric islands to pinned monolayers** The inertial sliding of submonolayer Xe islands on the Cu(111) surface just investigated offers an ideal playground to delve deeper into some important frictional phenomena. The long slip times and the theoretically demonstrated incommensurability between adsorbed and substrate lattices characterize the sliding as asymptotically superlubric for large island areas, limited only by defect- and edge-related friction. A sudden spontaneous compression upon monolayer completion caused by strong adhesion to the substrate is predicted and observed. Specific to Xe/Cu(111) is the ensuing $\sqrt{3}\times\sqrt{3}$ commensuration accidentally reached during compression, causing a peculiar transition from superlubric sliding with a large slip time to its sudden vanishing in the dense pinned state. Both the island superlubricity and the compressional transition, the latter generally leading from one to another incommensurate state, are general phenomena characterizing the sliding friction for adsorbates on crystalline surfaces, not specific to the system under study. The chapter of technological nanodesign addressing the control of crystal friction properties at the most intimate level will have to keep these elementary mechanisms into account.

## Acknowledgements

Work in Trieste was carried out under ERC Grant 320796 MODPHYSFRICT. Support from Regione Emilia Romagna, Project INTERMECH – MO.RE, SNSF Sinergia Contract CRSII2 136287/1 and EU COST Action MP1303 are also gratefully acknowledged.

## Author Contributions

G.M. conceived the project. G.P., A. di B. and S.V. deposited and characterized the Cu(111) samples. M.P., L.B and G.M. carried out the QCM experiments and analyzed the data. R.G., A.V. and E.T. developed the theoretical model and performed the numerical simulations. All authors discussed the results and contributed to the writing of the manuscript.

## Additional information

Supplementary information accompanies this paper at www.nature.com/naturenanotechnology. Reprints and permission information is available online at http://npg.nature.com/reprintsandpermissions/. Correspondence and requests for materials should be addressed to G.M.

## References

1. Socoliuc, A., Bennewitz, R., Gnecco, E. & Meyer, E. Transition from stick-slip to continuous sliding in atomic friction: Entering a new regime of ultralow friction. *Physical Review Letters* **92,** 134301 (2004).
2. Socoliuc, A. et al. Atomic-scale control of friction by actuation of nanometer-sized contacts. *Science* **313**, 207-210 (2006).
3. Krylov, S.Y. & Frenken, J.W.M. The physics of atomic-scale friction: Basic considerations and open questions. *Physica Status Solidi B-Basic Solid State Physics* **251**, 711-736 (2014).
4. Vanossi, A., Manini, N., Urbakh, M., Zapperi, S. & Tosatti, E. Colloquium: Modeling friction: From nanoscale to mesoscale. *Reviews of Modern Physics* **85**, 529-552 (2013).

5. Peyrard, M. & Aubry, S. Critical behaviour at the transition by breaking of analyticity in the discrete Frenkel-Kontorova model. *Journal of Physics C: Solid State Physics* **16**, 1593 (1983).
6. Hirano, M. & Shinjo, K. Atomistic locking and friction. *Physical Review B* **41**, 11837-11851 (1990).
7. Zhang, R.F. et al. Superlubricity in centimetres-long double-walled carbon nanotubes under ambient conditions. *Nature Nanotechnology* **8**, 912-916 (2013).
8. Nigues, A., Siria, A., Vincent, P., Poncharal, P. & Bocquet, L. Ultrahigh interlayer friction in multiwalled boron nitride nanotubes. *Nature Materials* **13**, 688-693 (2014).
9. Dienwiebel, M. et al. Superlubricity of graphite. *Physical Review Letters* **92**, 126101 (2004).
10. Filippov, A.E., Dienwiebel, M., Frenken, J.W.M., Klafter, J. & Urbakh, M. Torque and twist against superlubricity. *Physical Review Letters* **100**, 046102 (2008).
11. van Wijk, M.M., Dienwiebel, M., Frenken, J.W.M. & Fasolino, A. Superlubric to stick-slip sliding of incommensurate graphene flakes on graphite. *Physical Review B* **88**, 235423 (2013).
12. Dietzel, D. et al. Frictional duality observed during nanoparticle sliding. *Physical Review Letters* **101** (2008).
13. Dietzel, D., Feldmann, M., Schwarz, U.D., Fuchs, H. & Schirmeisen, A. Scaling Laws of Structural Lubricity. *Physical Review Letters* **111**, 235502 (2013).
14. Bohlein, T., Mikhael, J. & Bechinger, C. Observation of kinks and antikinks in colloidal monolayers driven across ordered surfaces. *Nature Materials* **11**, 126-130 (2012).
15. Vanossi, A., Manini, N. & Tosatti, E. Static and dynamic friction in sliding colloidal monolayers. *Proceedings of the National Academy of Sciences of the United States of America* **109**, 16429-16433 (2012).
16. Krim, J. Friction and energy dissipation mechanisms in adsorbed molecules and molecularly thin films. *Advances in Physics* **61**, 155-323 (2012).
17. Daly, C. & Krim, J. Sliding friction of solid xenon monolayers and bilayers on Ag(111). *Physical Review Letters* **76**, 803-806 (1996).
18. Coffey, T. & Krim, J. Impact of substrate corrugation on the sliding friction levels of adsorbed films. *Physical Review Letters* **95**, 076101 (2005).
19. Bruschi, L., Carlin, A. & Mistura, G. Depinning of atomically thin Kr films on gold. *Physical Review Letters* **88**, 046105 (2002).
20. Pierno, M. et al. Nanofriction of Neon Films on Superconducting Lead. *Physical Review Letters* **105**, 016102 (2010).
21. Bruschi, L. et al. Friction reduction of Ne monolayers on preplated metal surfaces. *Physical Review B* **81**, 115419 (2010).
22. Pierno, M. et al. Thermolubricity of gas monolayers on graphene. *Nanoscale* **6**, 8062-8067 (2014).
23. Cieplak, M., Smith, E.D. & Robbins, M.O. Molecular-origins of friction - the force on adsorbed layers. *Science* **265**, 1209-1212 (1994).
24. Tomassone, M.S., Sokoloff, J.B., Widom, A. & Krim, J. Dominance of phonon friction for a xenon film an a silver (111) surface. *Physical Review Letters* **79**, 4798-4801 (1997).
25. Righi, M.C. & Ferrario, M. Pressure induced friction collapse of rare gas boundary layers sliding over metal surfaces. *Physical Review Letters* **99**, 176101 (2007).
26. Righi, M.C. & Ferrario, M. Potential energy surface for rare gases adsorbed on Cu(111): parameterization of the gas/metal interaction potential. *Journal of Physics-Condensed Matter* **19**, 305008 (2007).
27. Varini, N. et al. Static friction scaling of physisorbed islands: the key is in the edge. *Nanoscale* **7**, 2093-2101 (2015).
28. Bruch, L.W., Diehl, R.D. & Venables, J.A. Progress in the measurement and modeling of physisorbed layers. *Reviews of Modern Physics* **79**, 1381-1454 (2007).

29. Thomy, A., Duval, X. & Regnier, J. Two-dimensional phase transitions as displayed by adsorption isotherms on graphite and other lamellar solids. *Surface Science Reports* **1**, 1-38 (1981).
30. Watts, E.T., Krim, J. & Widom, A. Experimental-observation of interfacial slippage at the boundary of molecularly thin-films with gold substrates. *Physical Review B* **41**, 3466-3472 (1990).
31. Persson, B.N. Sliding friction: physical principles and applications, Vol. 1. (Springer, 2000).
32. Seyller, T., Caragiu, M., Diehl, R.D., Kaukasoina, P. & Lindroos, M. Observation of top-site adsorption for Xe on Cu(111). *Chemical Physics Letters* **291**, 567-572 (1998).
33. Jupille, J., Ehrhardt, J.J., Fargues, D. & Cassuto, A. Study of xenon layers on a cu(111) surface. *Faraday Discussions* **89**, 323-328 (1990).
34. Braun, J., Fuhrmann, D., Siber, A., Gumhalter, B. & Woll, C. Observation of a zone-center gap in the longitudinal mode of an adsorbate overlayer: Xenon on Cu(111). *Physical Review Letters* **80**, 125-128 (1998).
35. Borovsky, B., Mason, B.L. & Krim, J. Scanning tunneling microscope measurements of the amplitude of vibration of a quartz crystal oscillator. *Journal of Applied Physics* **88**, 4017-4021 (2000).
36. Bruschi, L. & Mistura, G. Measurement of the friction of thin films by means of a quartz microbalance in the presence of a finite vapor pressure. *Physical Review B* **63**, 235411 (2001).
37. Pierno, M. et al. Nanofriction of adsorbed monolayers on superconducting lead. *Physical Review B* **84**, 035448 (2011).
38. Park, J.Y. et al. Adsorption and growth of Xe adlayers on the Cu(111) surface. *Physical Review B* **60**, 16934-16940 (1999).
39. Hasnain, J., Jungblut, S. & Dellago, C. Dynamic phases of colloidal monolayers sliding on commensurate substrates. *Soft Matter* **9**, 5867-5873 (2013).
40. Da Silva, J.L.F., Stampfl, C. & Scheffler, M. Xe adsorption on metal surfaces: First-principles investigations. *Physical Review B* **72**, 075424 (2005).
41. Da Silva, J.L.F., Ph.D. thesis.
42. Chesters, M., Hussain, M. & Pritchard, J. Xenon monolayer structures on copper and silver. *Surface Science* **35**, 161-171 (1973).